\documentclass[12pt]{article}
\usepackage{amsmath}
\usepackage{amsthm}
\usepackage{libertine}
\usepackage[utf8]{inputenc}
\usepackage[T1,T2A]{fontenc}
\usepackage[russian, english]{babel}
\usepackage[margin=1in]{geometry}
\usepackage{amsmath,amssymb} 
\usepackage{multicol}
\usepackage[shortlabels]{enumitem}
\usepackage{siunitx}
\usepackage{graphicx,url}
\usepackage{pgfplots}
\usepackage{listings}
\usepackage{tikz}
\usepackage{listings}
\usepackage{subcaption}
\usepackage{hyperref}
\usepackage[version=3]{mhchem} %%\textbf{IMPORTANT TO WRITE CHEMICALEQUATIONS}
\usepackage{listingsutf8}
\usepackage{comment}
\usepackage{authblk}

\pgfplotsset{width=10cm,compat=1.9}
\usepgfplotslibrary{external}
\newcommand{\bigzero}{\mbox{\normalfont\Large\bfseries 0}}
\newcommand{\rvline}{\hspace*{-\arraycolsep}\vline\hspace*{-\arraycolsep}}

\newcommand{\emp}[1]{{\bf  #1}}
\newcommand{\janos}[1]{{\bf  #1}}
\newcommand{\Zeta}{Z}
\newcommand{\Rb}{\mathbf{R}}
\newcommand{\rb}{\mathbf{r}}

\newcommand{\N}{\mathbb{N}}
\newcommand{\R}{\mathbb{R}}

\newcommand{\dsqr}{b^2-c^2-4ad+2eA}
\newcommand{\deaF}{-\frac{ae}{2}-2bC+dA}

\renewcommand{\Re}[1]{\mathfrak{Re}[#1]}
\renewcommand{\Im}[1]{\mathfrak{Im}[#1]}
\theoremstyle{definition}
\newtheorem{lemma}{Lemma}
\newtheorem{theorem}{Theorem}
\newtheorem{proposition}{Proposition}

\newtheorem{remark}{Remark}

\newcommand{\Pf}[1]{{\textit{Proof.}\ }#1}

\newcommand{\res}[2]{\pm\left(\frac{\sqrt{\sqrt{#1^2+#2^2}+#1}}{\sqrt{2}}
+i\frac{#2}{\sqrt{2}\sqrt{\sqrt{#1^2+#2^2}+#1}}\right)}
\newcommand{\result}[2]{\pm\left(\frac{#2}{\sqrt{2}\sqrt{\sqrt{#1^2+#2^2}-#1}}
+i\frac{\sqrt{\sqrt{#1^2+#2^2}-#1}}{\sqrt{2}}\right)}
\newcommand{\ratio}[4]{\frac{{#1#2+#3#4}}{#2^2+#4^2}+i\frac{-#1#4+#3#2}{#2^2+#4^2}}
\title{Symbolic solution of systems of polynomial differential equations 
via the Cauchy--Riemann equation. Applications to kinetic differential equations}
\author[1]{Kelvin Kiprono}
\author[1,2]{J\'anos T\'oth}
\affil[1]
{Department of Analysis and Operations Research,
Institute of Mathematics,
Budapest University of Technology and Economics,
Műegyetem rkp. 3., H-1111 Budapest, Hungary}
\affil[2]{
Chemical Kinetics Laboratory,
Institute of Chemistry, 
ELTE E\"{o}tv\"{o}s Lor\'{a}nd University, 
P\'azm\'any P. s\'et\'any 1/A, H-1117 Budapest}
\begin{document}
\maketitle
%%%%%%%%%%%%%%%%%%%%%%%%%%%%%%%%%%%%%%%%%%%
\begin{abstract}
The differential equations of chemical kinetics are systems of nonlinear (polynomial) differential equations, therefore their solutions cannot usually be found in symbolic form. Here we offer a method to solve classes of kinetic differential equations based on the Cauchy--Riemann equations. It turns out that the method can be used to symbolically solve some polynomial differential equations that are not necessarily kinetic, as well. 
\end{abstract}

%%%%%%%%%%%%%%%%%%%%%%%%%%%%%%%%%%%%%%%%%%%
\setcounter{tocdepth}{4}
\tableofcontents
%%%%%%%%%%%%%%%%%%%%%%%%%%%%%%%%%%%%%%%%%%%
\section{Introduction}
%%%%%%%%%%%%%%%%%%%%%%%%%%%%%%%%%%%%%%%%%%%
The solutions of the differential equations of chemical reaction kinetics generally cannot be explicitly given symbolically, except in a few, very special cases. 
Here we consider the solution of a kinetic differential equation
as explicit or symbolic, if the concentrations of the species are given as a function of time, not necessarily via elementary functions, and infinite series also including. 
We categorize solutions into this class also if the inverse function of a concentration vs. time function is given. (See a review on closed form solutions  \cite{borweincrandall}.)
It is almost as useful as these if the concentration of one species is given as the function of the concentration of another. As is customary, we define a solution as symbolic if a finite algorithm (typically consisting of fewer than 10 steps in practice) is provided, allowing for the construction of the solution, as seen in the derivation of the solution of cubic or similar equations, or simple differential equations. However, it is important to note that this definition may imply that solutions with symbolic parameters (or with "unlucky" numeric values) cannot be practically calculated in a usable form.

Preliminary work in this field includes \cite{szabo} presenting a large collection of explicitly solvable cases. 
A major merit of that paper is that the author also attached a chemical example with references to all the cases he was able to solve. 
In some cases (as it is known from the theory of differential equations) he was only able to provide the inverse of a concentration vs. time function.
Also, it happened sometimes that he could give the \emph{selectivity curves}, i.e. the concentration of one species as the function of the concentration of another. These are the projections of the trajectories in mathematical terms.

Another large collection has been published in \cite{rodiginrodigina}. These authors have shown how to solve the case when all the reaction steps have an order not larger than one: all the reaction steps are of order one or zero. No wonder that one of their tools is Laplace transformation.
Although the solution of linear differential equations that can be given is generally considered to be explicit, it is only relatively explicit, so far as the eigenvalues and eigenvectors are known---which is not the case if the number of variables is larger than four.

Recently, Lente 
\cite{lenteanalytical,lentedeterministic,lentekinetics,lentemixed}
solved quite a few kinetic differential equations: those describing reactions with two steps and not being of the order larger than two.
The explicit solutions contain special functions like the hypergeometric function, exponential integral, etc., and also infinite series. 
Again, in some cases, he could give the \emph{selectivity curves}. 
The work by Park \cite{park} has similar, but more limited ambitions. Hernandez and Lubenia \cite{hernandezlubenia} could provide a symbolic representation of the stationary states of kinetic differential equations.

Kamke has gathered the classical collection of almost 2000 symbolically solvable ordinary differential equations \cite{Kamke}. (He also published a volume on partial differential equations: \cite{kamkeII}.) These days a minimal requirement for a mathematical program package, like e.g. the Wolfram Language, is to be capable of reproducing Kamke's solutions and going beyond. A more recent and more modern collection can be found on the Internet: \cite{polyanin}.
Still, this is not the end of the story because papers studying first integrals \cite{nagytothquadratic} or symmetries \cite{gineromanovski,ginevalls} also contribute to this area. Let us note that in pure mathematics, finding explicit solutions to ordinary differential equations is a separate area, papers belonging to this are classified under 34A05 of AMS MSC, see e.g. the works by Calogero and his coworkers \cite{calogeroconteleyvraz,calogeropayandehcoupled,calogeropayandehODE,calogeropayandehperiodic}. 
As our work goes beyond the boundaries of reaction kinetics, the present results are an addition to this kind of Painlevé approach as well, as \cite[p. 1]{calogeroconteleyvraz} puts it.

It is quite common in the literature of reaction kinetics that a special, individual equation (not a parameterized class) is solved, we are not interested in these cases as these may count to millions.

In the present paper, we aim to find kinetic differential equations of the Cauchy--Riemann--Erugin type
(cf. \cite[V. \S 2, $4^{o}$]{matveev}, and \cite[Ex. 1091., 1092.]{matveev}, furthermore \cite[p. 116]{bogdanovsyroid},\cite{totherdisymmetry}): two-variable equations with a right-hand side the components of which can be considered to be the real and imaginary part of a differentiable complex function.  
Then, we are going to solve the kinetic differential equations belonging to this class. 
Next, we shall give a few (second and third-order) complex chemical reactions whose induced kinetic differential equation can be solved by the method.
To find a realization of a given kinetic differential equation is far from being a trivial problem, see e.g. \cite{craciunjinyu,craciunjohnstonszederkenyitonellotothyu,deshpande,harstoth,johnstonsiegelszederkenyi}.

A restriction of our method is that the reaction rate coefficients should fulfill some equalities, this is not quite uncommon either in reaction kinetics or in the theory of differential equations.
The method is extended to four-variable equations.
Although kinetic differential equations form a special subclass of polynomial differential equations, our method will only be easier to apply to them, the
results of the method extend beyond this field.

If our equation is not of the Cauchy--Riemann--Erugin type but is close to that then the solutions will also be close. However, the trajectories may suffer quantitative or even qualitative distortions.

The structure of our paper is as follows. 
Section \ref{sec:method} presents the essence of our method. 
Section \ref{sec:examples} starts with a trivial example: the equation of first-order reactions in two variables. 
The next part is about second and third-order reactions, providing more useful results from the point of view of kinetics because one has more free parameters in higher-order cases. 
This section contains a nice compact form of the original real equations and the complex equation as well (Theorem \ref{thm:main}) that may be the result most interesting from the mathematical point of view.
Section \ref{sec:examples} applies the multivariate Cauchy--Riemann equation to extend the method to four (or, \(2n\)) variable cases. 
Then, perspectives of approximations obtained by the method, partial differential equations, and possible applications of quaternions are also treated or at least mentioned in the Discussion and Outlook section. In all cases, we start without the restriction to kinetic differential equations
\cite{feinbergbook,tothnagypapp}, the results will only be specialized at the end. 
Most of the general and special calculations are relegated to the Appendix.
%%%%%%%%%%%%%%%%%%%%%%%%%%%%%%%%%%%%%%%%%%%%
\section{The method}\label{sec:method}
%%%%%%%%%%%%%%%%%%%%%%%%%%%%%%%%%%%%%%%%%%%%
The steps of the method are as follows.
\begin{enumerate}
\item 
Start from a two-variable differential equation with a polynomial right-hand side
\begin{equation}\label{eq:twovariable}
\dot{x}=u\circ(x,y),\quad \dot{y}=v\circ(x,y),
\end{equation}
and apply the restrictions on the coefficients coming from the assumption that the right-hand sides are the real and imaginary parts of a differentiable complex function \(f,\) i.e. they fulfill the Cauchy--Riemann conditions:
\begin{equation}\label{eq:cauchy}
\partial_1u=\partial_2v,\quad\partial_2u=-\partial_1v.    
\end{equation}
\item 
Rewrite the differential equations into a single equation for the complex function \(f.\)
\item 
Solve the equation for \(z:=x+iy.\)
\item 
Determine the real and imaginary parts of the solution.
\item 
Determine the general form of the induced kinetic differential equation of two-species first-order reactions, taking care of the absence of negative cross-effect, and specialize the above-obtained results to this special case.
\item 
Finally, try to find "simple" realizations for the pair of real differential equations: with as small many complexes and reaction steps as possible. 
Note that this is not our final goal here.
\end{enumerate}
Let us mention in passing that the relations \eqref{eq:cauchy} can be rephrased as
\begin{equation}
f'J=Jf'\text{ with }
J:=
\begin{bmatrix}
0&1\\-1&0    
\end{bmatrix}.
\end{equation}
For the multivariate case of two complex functions and two complex variables such as the one below.
\begin{align*}
    f_1(z_1,z_2) &= u_1(x_1,x_2,y_1,y_2)+iv_1(x_1,x_2,y_1,y_2) \\
    f_2(z_1,z_2) &= u_2(x_1,x_2,y_1,y_2)+iv_2(x_1,x_2,y_1,y_2)
\end{align*}
we have:
\begin{equation*}
f'J=Jf'\text{ with }
J:=
\begin{pmatrix}
  \bigzero & \rvline
 & \begin{matrix}
  1 & 0 & 0 & 0 \\
  0 & 1 & 0 & 0 \\
  0 & 0 & 1 & 0 \\
  0 & 0 & 0 & 1
 \end{matrix}\\
\hline
\begin{matrix}
 -1 & 0 & 0 & 0 \\
 0 & -1 & 0 & 0 \\
 0 & 0 & -1 & 0 \\
 0 & 0 & 0 & -1
 \end{matrix}

& \rvline & \bigzero \\
 \end{pmatrix}
 \end{equation*}
In general, if we have \(n\) multivariate complex functions with \(n\) complex variables, then its Jacobian matrix is given by the following block matrix:
\begin{equation}
    J :=
\left[
\begin{array}{c|c}
\bigzero & \mathbf{I_{2n}} \\
\hline
-\mathbf{I_{2n}} & \bigzero
\end{array}
\right]
\end{equation}
%%%%%%%%%%%%%%%%%%%%%%%%%%%%%%%%%%%%%%%%%%%%
\section{Special cases}\label{sec:examples}
%%%%%%%%%%%%%%%%%%%%%%%%%%%%%%%%%%%%%%%%%%%%
We treat in some detail cases that are either relevant from the point of view of reaction kinetics or are not too complicated allowing us to carry out the calculations symbolically up to the end.
%%%%%%%%%%%%%%%%%%%%%%%%%%%%%%%%%%%%%%%%%%%%
\subsection{Linear systems, first-order reactions}
%%%%%%%%%%%%%%%%%%%%%%%%%%%%%%%%%%%%%%%%%%%%
Let us apply the method in a case when it is not needed, but transparently shows the essential steps.
\begin{enumerate}
\item 
The general form of two-variable linear differential equations is:
\begin{equation*}
    \dot{x} = a+bx+cy,\quad
    \dot{y} = A+Bx+Cy
\end{equation*}
with the real coefficients \(a,b,c,A,B,C.\)
Assuming that the right-hand sides of the above differential equations are real and imaginary parts of the differentiable complex function \(f,\) i.e. they are satisfying the Cauchy-Riemann conditions, we get \(C=b, B=-c,\) or
\begin{equation*}
    \dot{x}=a+bx+cy,\quad
    \dot{y}=A-cx+by.\label{eq:linearCR}
\end{equation*}
Without excluding relevant interesting cases we assume \(b^2+c^2\neq0.\)
In this very special (linear) case, one can solve the system of equations \eqref{eq:linearCR} with the initial condition 
\begin{equation}\label{eq:inicond}
x(0)=x_0,\quad y(0)=y_0
\end{equation} 
as
\begin{align}
x(t)&=\frac{e^{b t}(\cos (c t)X+\sin (c t)Y)-a b+A c}{b^2+c^2},\label{eq:linearsolx}\\
y(t)&=\frac{e^{b t} (\cos (c t) Y-\sin (c t) X)-ac-Ab}{b^2+c^2},\label{eq:linearsoly}
\end{align}
with
\begin{equation}
X:=a b-A c+x_0(b^2+c^2),\quad
Y:=a c+A b+y_0 (b^2+c^2).
\end{equation}
We shall not be as lucky as this later.
\item 
The equation for \(z:=x+iy\) 
if \(x\) and \(y\) obeys \eqref{eq:linearCR} and
\eqref{eq:inicond} is as follows:
\begin{equation}\label{eq:lincomplex}
\dot{z}=a+iA+(b-ic)z, z(0)=x_0+iy_0.
\end{equation}
\item
The solution of \eqref{eq:lincomplex} can easily be found such as
\begin{equation}\label{eq:lincomplexsol}
z(t)=e^{(b-ic)t}\left(z(0)+\frac{a+iA}{b-ic}\right)-\frac{a+iA}{b-ic},
\end{equation}
\item that can be rewritten using Euler's formula and Lemma \ref{lem:formulas} as
\begin{align}\label{eq:linearcomplexsolexpanded}
x(t)&=e^{bt}\left(\cos(bt)(x_0+\frac{ab-Ac}{b^2+c^2})+\sin(bt)(y_0+\frac{ac+Ab}{b^2+c^2})\right)-\frac{ab-Ac}{b^2+c^2},\\
y(t)&==e^{bt}\left(\sin(bt)(x_0+\frac{ab-Ac}{b^2+c^2})+\cos(bt)(y_0+\frac{ac+Ab}{b^2+c^2})\right)-\frac{ac+Ab}{b^2+c^2}
\end{align}
which is the same as given in \eqref{eq:linearsolx}--\eqref{eq:linearsoly}.

\item
To have a kinetic (or, Hungarian) polynomial differential equation no negative cross-effect (\cite{harstoth}) can be present, thus we have the restrictions \(a, c, A, B\ge0.\) 
These, together with the Cauchy--Riemann conditions imply that
\(c=0.\) Again, we are excluding here the trivial case \(b=0.\)
Specializing \eqref{eq:linearsolx}--\eqref{eq:linearsoly} to this case gives
\begin{align}
x(t)&=(x_0+\frac{a}{b})e^{bt}-\frac{a}{b}, \label{eq:linearkineticsol1}\\
y(t)&=(y_0+\frac{A}{b})e^{bt}-\frac{A}{b}.\quad (t\in\R).\label{eq:linearkineticsol2}
\end{align}
\item 
Now we find a realization of the differential equation (of the one that is not only of the Cauchy--Riemann type but also kinetic)
\begin{equation}
    \dot{x}=a+bx\quad
    \dot{y}=A+by\label{eq:linearCRkinetic}
\end{equation}
by finding a reaction inducing it.
A solution with a minimal number of complexes and reaction steps is
seen in Fig. \ref{fig:firstorderFHJneg} when
\(b<0.\)
\begin{figure}[ht]
  \centering
  \includegraphics[width=0.7\textwidth]{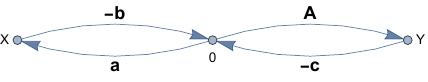}
  \caption{A reaction with a minimal number of complexes and reaction steps inducing the kinetic differential equation \eqref{eq:linearCRkinetic} with \(b<0\).}
  \label{fig:firstorderFHJneg}
\end{figure}
When  \(b>0, \) we have the reaction in Fig. \ref{fig:firstorderFHJpos}
\begin{figure}[ht]
  \centering
  \includegraphics[width=0.7\textwidth]{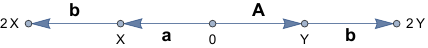}
  \caption{A reaction with a minimal number of complexes and reaction steps inducing the kinetic differential equation \eqref{eq:linearCRkinetic} with \(b>0\).}
  \label{fig:firstorderFHJpos}
\end{figure}
\end{enumerate}
%%%%%%%%%%%%%%%%%%%%%%%%%%%%%%%%%%%%%%%%%%%%
\subsection{Second-degree systems, second-order reactions}
%%%%%%%%%%%%%%%%%%%%%%%%%%%%%%%%%%%%%%%%%%%%
Let us apply the method.
\begin{enumerate}
\item
The general form of two-variable differential equations with a second-degree right-hand side is:
\begin{equation*}
    \dot{x} = a+bx+cy+dx^2+exy+fy^2,\quad
    \dot{y} = A+Bx+Cy+Dx^2+Exy+Fy^2.
\end{equation*}
Assuming that components of the right-hand side of the above differential equation form the real and imaginary parts of a differentiable complex function \(f\) satisfying the Cauchy--Riemann conditions, we have:
\begin{equation*}
b+2dx+ey=C+Ex+2Fy, \quad c+ex+2fy = -B-2Dx-Ey.
\end{equation*}
Equating the coefficients on the right--hand side and those of the left hand--side in each of  the equations above we get:
\[
B=-c, C=b, D=-\frac{e}{2}, E=2d=-2f, F=\frac{e}{2}.
\]
Since \(E=2d\) and \(E=-2f\) it follows that \(f=-d,\) thus we can eliminate \(f.\)
We obtain the following differential equation:
\begin{align}
\dot{x} =a+bx+cy+dx^2+exy-dy^2,\label{eq:realsecondx}\\
\dot{y} =A-cx+by-\frac{e}{2}x^2+2dxy+\frac{e}{2}y^2.\label{eq:realsecondy}
\end{align}
Now it does not seem possible to solve this system with classical methods.
\item 
However, the equation for \(z:=x+iy\) 
if \(x\) and \(y\) obeys \eqref{eq:realsecondx}--\eqref{eq:realsecondy} and
\eqref{eq:inicond} is as follows:
\begin{equation}\label{eq:secondcomplex}
\dot{z}=a+iA+(b-ic)z+(d-i\frac{e}{2})z^2,\quad z(0)=x_0+iy_0.
\end{equation}
\item 
The solution of \eqref{eq:secondcomplex} can be found, see the Appendix, Subsection \ref{subsec:addsecond}.
\item 
The final result giving the solution of the real system of equations \eqref{eq:realsecondx}--\eqref{eq:realsecondy} together with the initial condition \eqref{eq:inicond} can be given, but it is so complicated that we only give it in special cases.
\item
Because of the absence of a negative cross-effect, we have the restrictions 
\(a,A,\ge0, d,e\le0\) and \(c=0.\) 
Specializing the general solution to this case gives the solution of the kinetic differential equation.
\item
Now we find a realization of the differential equation (of the one that is not only of the Cauchy--Riemann type but also kinetic)
\begin{equation}\label{eq:secondkinetic}
\dot{x} =a+bx+dx^2+exy-dy^2,\quad
\dot{y} =A+by-\frac{e}{2}x^2+2dxy+\frac{e}{2}y^2+dy.
\end{equation}
with \(a,A\ge0; d,e\le0, \) and \(b<0, b=0, b>0.\)

A relatively simple realization is the canonic realization, see Fig. \ref{fig:secondreal}. 
\begin{figure}[ht]
  \centering
  \includegraphics[width=0.3\textwidth]{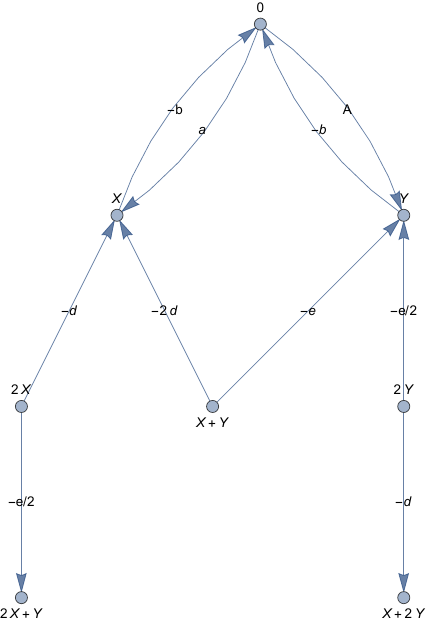}
  \includegraphics[width=0.3\textwidth]{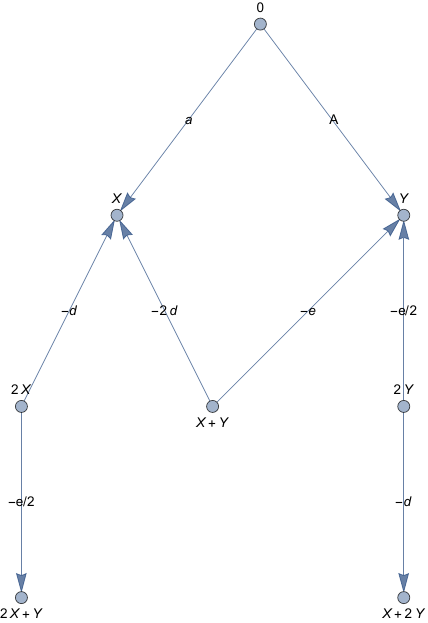}
  \includegraphics[width=0.3\textwidth]{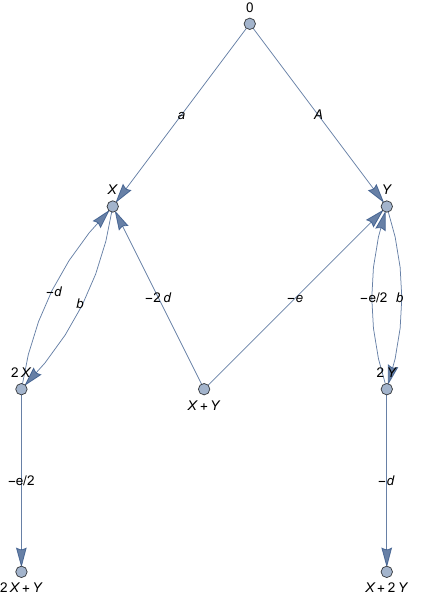}
  \caption{The canonic realization of Eq. \eqref{eq:secondkinetic}
  for $b<0,b=0,b>0$ respectively.}
  \label{fig:secondreal}
\end{figure}

\end{enumerate}
Let us make a few remarks on the realization problem.
The right-hand sides of the equation \eqref{eq:secondkinetic}
are sums of monomials with the exponents 
\begin{equation*}
\begin{bmatrix}0\\0\end{bmatrix},
\begin{bmatrix}1\\0\end{bmatrix}, 
\begin{bmatrix}0\\1\end{bmatrix}, 
\begin{bmatrix}2\\0\end{bmatrix}, 
\begin{bmatrix}1\\1\end{bmatrix}, 
\begin{bmatrix}0\\2\end{bmatrix}. 
\end{equation*}
The complexes corresponding to these vectors, i.e.
\begin{equation*}
\ce{0}, \ce{X}, \ce{Y}, \ce{2X}, \ce{X + Y}, \ce{2Y}   
\end{equation*}
will surely play a role as a reactant complex. 
As to the realizations, we may have different goals:
\begin{enumerate}
\item 
No more complexes be used.
\item
Let the number of reaction steps be as small as possible.
\item
Let the product complexes (if we are forced to add new ones) have integer coefficients.
\item
Let the realization be reversible, 
weakly reversible, 
complex balanced, 
detailed balanced,
let it have as small a deficiency as possible.
\end{enumerate}
There is abundant literature on the topics from \cite{harstoth} through \cite{craciunjohnstonszederkenyitonellotothyu} until recent papers. 
Here we shall be content with the canonic realizations above, because
they are not too bad in the sense that they only use two additional complexes beyond the absolutely necessary ones.
%%%%%%%%%%%%%%%%%%%%%%%%%%%%%%%%%%%%%%%%%%%%
\subsection{Third-degree systems, third-order reactions}
%%%%%%%%%%%%%%%%%%%%%%%%%%%%%%%%%%%%%%%%%%%%
As third-order reactions are not always excluded, it is worth studying this case, as well. Furthermore, the number of free (adjustable) parameters is much larger here, therefore the explicit solutions can be used for many reactions.
\begin{enumerate}
\item
The general form of two-variable differential equations with a third-degree right-hand side is:
\begin{align*}
\dot{x}&=a+bx+cy+dx^2+exy+fy^2+gx^3+hx^2y+jxy^2+ky^3,\\
\dot{y}&=A+Bx+Cy+Dx^2+Exy+Fy^2+Gx^3+Hx^2y+Jxy^2+Ky^3.
\end{align*} 
The Cauchy--Riemann conditions mean here:
\begin{align}
b+2dx+ey+3gx^2+2hxy+jy^2&=C+Ex+2Fy+Hx^2+2Jxy+3Ky^2\\
c+ex+2fy+hx^2+2jxy+3ky^2&=-(B+2Dx+Ey+3Gx^2+2Hxy+Jy^2).    
\end{align}
Comparing the coefficients on the left-hand side and right-hand side of both equations we obtain the following
results on coefficients:
\begin{align}
B=-c, C=b, D=-\frac{e}{2}, E=2d, E=-2f, F=\frac{e}{2},\\
 G=-\frac{h}{3}, H=3g, H=-j, J=h, J=-3k, K=\frac{j}{3}.
\end{align}

Since 
\(E=2d\) and \(E=-2f\) it follows that \(f=-d,\) thus we can eliminate \(f.\)
Similarly, 
\(H=3g\) and \(H=-j\) imply \(j=-3g;\)
\(J=h\) and \(J=-3k\) imply \(k=-\frac{h}{3};\)
therefore
\begin{align}
f&=-d, j=-3g, k=-\frac{h}{3};\\
B&=-c, C=b, D=-\frac{e}{2}, E=2d, F=\frac{e}{2},\\
G&=-\frac{h}{3}, H=3g, J=h, K=-g.
\end{align}
We obtain the following differential equation:
\begin{align}
\dot{x}&=a+bx+cy+dx^2+exy-dy^2+gx^3+hx^2y-3gxy^2-\frac{h}{3}y^3,\\
\dot{y}&=A-cx+by-\frac{e}{2}x^2+2dxy+\frac{e}{2}y^2-\frac{h}{3}x^3+3gx^2y+hxy^2-gy^3.
\label{eq:thirdCR}
\end{align}
Again, it does not seem possible to solve this system with classical methods.
\item 
However, the equation for \(z:=x+iy\) 
if \(x\) and \(y\) obeys \eqref{eq:thirdCR} and
\eqref{eq:inicond} is as follows:
\begin{equation}\label{eq:thirdcomplex}
\dot{z}= (h-i\frac{g}{3})z^3+(d-i\frac{e}{2})z^2+(b-ic)z+(a+iA),\quad z(0)=x_0+iy_0.
\end{equation}
\item
The implicit solution of \eqref{eq:thirdcomplex} can be found such as
\begin{align}\label{eq:thirdcomplexsol}
&(z(t)-z_1)(z(t)-z_2)(z(t)-z_3)\\
&=(z(0)-z_1)(z(0)-z_2)(z(0)-z_3)\exp{(z_1-z_2)(z_1-z_3)(z_2-z_3)(a-i\frac{b}{3})t},
\end{align}
where \(z_1, z_2\) and \(z_3\) are the roots of the polynomial 
\begin{equation*}
(z\mapsto  (h-i\frac{g}{3})z^3+(d-i\frac{e}{2})z^2+(b-ic)z+(a+iA),
\end{equation*}
and \(z(0):=x_0+iy_0.\)
To get an explicit solution, the application of the Cardano formula is needed, and
the result would not be really useful. 
Even the specialization to the kinetic case is not simpler.
To show the steps of the process, we do all the calculations in Subsection \ref{subsec:thirddegreecase} of the Appendix for a few cases.
\item 
Now we find a realization of the differential equation of the one that is not only of the Cauchy--Riemann type but also kinetic. The above equation is kinetic if \(c=0; a,A\ge0;d,h,e\le0.\) However, there are no restrictions to the signs of \(b\) and \(g.\)
\begin{align}
\dot{x}&=a+bx+dx^2+exy-dy^2+gx^3+hx^2y-3gxy^2-\frac{h}{3}y^3,\\
\dot{y}&=A+by-\frac{e}{2}x^2+2dxy+\frac{e}{2}y^2-\frac{h}{3}x^3+3gx^2y+hxy^2-gy^3.
\label{eq:thirdCRkinetic}
\end{align}
 We shall treat the case
 \(b<0\) and \(g<0\) in detail and leave the three remaining cases  
 \(b\le0, g\ge0\), and
\(b\ge0, g\le0\), and
\(b\ge0, g\ge0\), and also the cases when some of the constants are zero, to the reader.
A relatively simple realization is the canonic realization: 
Note that beyond the complexes corresponding to the exponents of the monomials on the right-hand sides we needed four more complexes:
\ce{2X + 2Y},\ce{X + 3 Y}, \ce{3 X + Y}, and \ce{4Y}.
Fig. \ref{fig:third-order} shows the FHJ graph for the canonic realization of the Eq. \eqref{eq:thirdCRkinetic}.
\begin{figure}[ht]
  \centering
    \includegraphics[width=0.8\textwidth]{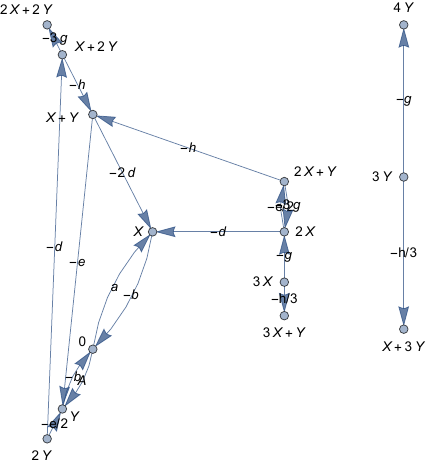}
    \caption{Canonic realization in the third order case when \(b<0,g<0\).}
    \label{fig:third-order}
\end{figure}
\end{enumerate}
Finally, we mention that it is mathematically possible to apply our method in the fourth-order case, but it is not worth the price. 
%%%%%%%%%%%%%%%%%%%%%%%%%%%%%%%%%%%%%%%%%%%%
\subsection{\textit{R}th-degree systems}
%%%%%%%%%%%%%%%%%%%%%%%%%%%%%%%%%%%%%%%%%%%%
The general case is interesting from the point of view of differential equations, certainly not from the point of view of formal reaction kinetics. Some of the earlier results can be summarized and generalized in a nice statement.
\begin{theorem}\label{thm:main}
Let \(R\in\N_0,\) and suppose that the polynomials \(u,v\)  defined as
\begin{align*}
u(x,y)&:=\sum_{r=0}^{R}\sum_{s=0}^{r}a_r^sx^{r-s}y^s,\quad
v(x,y):=\sum_{r=0}^{R}\sum_{s=0}^{r}A_r^sx^{r-s}y^s\\
&(a_r^s, A_r^s\in\R,s=0,1,2,\dots,r; r=1,2,\dots,R)
\end{align*} 
satisfy the Cauchy--Riemann equations \eqref{eq:cauchy}.
Then, the system of differential equations
\begin{equation}\label{eq:system}
\dot{x}=u\circ(x,y),\quad\dot{y}=v\circ(x,y)
\end{equation}
can be simplified to
\begin{align}
\dot{x}&=a^0_0+\sum_{r=1}^{R}\left(a^0_r\Re{(x+iy)^r}+\frac{a^1_r}{r}\Im{(x+iy)^r}\right),\\
\dot{y}&=A^0_0+\sum_{r=1}^{R}\left(a^0_r\Im{(x+iy)^r}-\frac{a^1_r}{r}\Re{(x+iy)^r}\right)
\end{align}
Furthermore, for the complex-valued function \(z:=x+i y\) one has
\begin{equation*}
\dot{z}=a^0_0+iA^0_0+\sum_{r=1}^{R}(a_r^0-i\frac{a_r^{1}}{r})z^r.
\end{equation*}
\end{theorem}
\begin{remark}
Note that the number of independent parameters left is \(2R+1,\) showing that the Cauchy--Riemann equations mean strong restrictions for the relations between the parameters. Furthermore, the coefficients of the terms with different degrees are independent, therefore it is enough to do the calculations with a single term.   
\end{remark}
\Pf{
\begin{enumerate}
\item
A direct comparison of the coefficients in the Cauchy--Riemann equations gives
\begin{align}
A^s_r&=\frac{r-s+1}{s}a^{s-1}_r\quad&(s=1,2,3,\dots,r; r=1,2,3,\dots,R),\label{eq:direct1}\\
A^s_r&=-\frac{s+1}{r-s}a^{s+1}_r\quad&(s=0,1,2,\dots,r-1; r=1,2,3,\dots,R).\label{eq:direct2}
\end{align}
Eqs.\eqref{eq:direct1} and \eqref{eq:direct2} imply
\begin{align}\label{eq:indirect1}
a^s_r=-\frac{(r-s+2)(r-s+1)}{s(s-1)}a^{s-2}_r\quad
(s=2,3,\dots,r; r=2,3,\dots,R).
\end{align}
The recursive application of \eqref{eq:indirect1}  (with \(i:=\sqrt{-1}\)) leads to
\begin{align}
a^s_r&=i^s\binom{r}{s}a^0_r,\quad &(s=0,2,4,6,\dots; r=2,3,\dots,R);\label{eq:recxeven}\\
a^s_r&=i^{s-1}\frac{1}{r}\binom{r}{s}a^1_r,\quad &(s=1,3,5,\dots; r=2,3,\dots,R).\label{eq:recxodd}
\end{align}
Therefore the \(r\mathrm{th}\) term of the first equation of the system \eqref{eq:system} can be written in the following ways:
\begin{align}
&
i^0\binom{r}{0}a^0_rx^r+
i^0\frac{1}{r}\binom{r}{1}a^1_rx^{r-1}y+
i^2\binom{r}{2}a^0_rx^{r-2}y^2+
i^2\frac{1}{r}\binom{r}{3}a^1_rx^{r-3}y^3\\
&+i^4\binom{r}{4}a^0_rx^{r-4}y^4+
i^4\frac{1}{r}\binom{r}{5}a^1_rx^{r-5}y^5+\dots\\
&=a^0_r\left(i^0\binom{r}{0}x^r+i^2\binom{r}{2}x^{r-2}y^2+i^4\binom{r}{4}x^{r-4}y^4+\dots\right)\\
&-i\frac{a^1_r}{r}\left(i^1\binom{r}{1}x^{r-1}y+i^3\binom{r}{3}x^{r-3}y^3+i^5\binom{r}{5}x^{r-5}y^5+\dots\right)\\
&=a^0_r\Re{(x+iy)^r}+\frac{a^1_r}{r}\Im{(x+iy)^r}.
\end{align}
\item
Similarly, the recursive application of \eqref{eq:direct1} leads to
\begin{align}
A^s_r&=i^{s-2}\frac{1}{r}\binom{r}{s}a^1_r,\quad (s=2,4,6,\dots; r=2,3,\dots,R;)\label{eq:recyeven}\\
A^s_r&=i^{s-1}\binom{r}{s}a^0_r,\quad (s=1,3,5,\dots; r=2,3,\dots,R.)\label{eq:recyodd}
\end{align}
Therefore the \(r\mathrm{th}\) term of the second equation of the system \eqref{eq:system} can be written in the following ways:
\begin{align}
&
-\frac{1}{r}\binom{r}{0}a^1_r x^r+
\binom{r}{1}a^0_r x^{r-1}y
+\frac{1}{r}\binom{r}{2}a^1_r x^{r-2}y^2
-\binom{r}{3}a^0_rx^{r-3}y^3\\
&-\frac{1}{r}\binom{r}{4}a^1_rx^{r-4}y^4+
\binom{r}{5}a^0_rx^{r-5}y^5+\dots\\
&=a^0_r\left(
i^0\binom{r}{1}x^{r-1}y+i^2\binom{r}{3}x^{r-3}y^3+i^4\binom{r}{5}x^{r-5}y^5+\dots\right)\\
&-
\frac{a^1_r}{r}\left(i^0\binom{r}{0}x^{r}+i^2\binom{r}{2}x^{r-2}y^2+i^4\binom{r}{4}x^{r-4}y^4+\dots\right)\\
&=a^0_r\Im{(x+iy)^r}-\frac{a^1_r}{r}\Re{(x+iy)^r}.
\end{align}
\item  
Finally, the equation for \(z:=x+iy\) 
if \(x\) and \(y\) obeys \eqref{eq:cauchy} and
\eqref{eq:inicond} is as follows:
\begin{align}
\dot{z}&=a^0_r\Re{(x+iy)^r}-i\frac{a^1_r}{r}\Im{(x+iy)^r}+i(a^0_r\Im{(x+iy)^r}+\frac{a^1_r}{r}\Re{(x+iy)^r})\label{eq:nthcomplex}\\
&=\sum_{r=0}^{R}(a_r^0-\frac{a_r^{1}}{r})z^r ,\quad z(0)=x_0+iy_0.
\end{align}
\end{enumerate}
}
%%%%%%%%%%%%%%%%%%%%%%%%%%%%%%%%%%%%%%%%%%%%%%%%%%%%%%%%%%%%%%
\section{Application of multivariate complex functions}\label{sec:multi}
%%%%%%%%%%%%%%%%%%%%%%%%%%%%%%%%%%%%%%%%%%%%%%%%%%%%%%%%%%%%%%%%%%
Here we extend our method to treat differential equations in multiple variables
\cite{ronkin,wirtinger}.
Suppose, \(u\) and \(v\) are the real and imaginary parts of the multivariate complex function \(f.\) 
Then, the Cauchy--Riemann equations are as follows.
\begin{equation}
\frac{\partial u}{\partial x_j} = \frac{\partial v}{\partial y_j},\quad
\frac{\partial u}{\partial y_j} =-\frac{\partial v}{\partial x_j}.
\end{equation}
Consider more specifically the case when we have two complex functions and two complex variables such as 
\begin{align*}
f_1(z_1,z_2)&=u_1(x_1,x_2,y_1,y_2)+iv_1(x_1,x_2,y_1,y_2), \\
f_2(z_1,z_2)&=u_2(x_1,x_2,y_1,y_2)+iv_2(x_1,x_2,y_1,y_2),
\end{align*}
then the equalities are:
\begin{align}
&\frac{\partial u_1}{\partial x_1} = \frac{\partial v_1}{\partial y_1},\quad
\frac{\partial u_1}{\partial y_1} =-\frac{\partial v_1}{\partial x_1},\quad
\frac{\partial u_1}{\partial x_2} = \frac{\partial v_1}{\partial y_2},\quad
\frac{\partial u_1}{\partial y_2} =-\frac{\partial v_1}{\partial x_2},\\\\
&\frac{\partial u_2}{\partial x_1} = \frac{\partial v_2}{\partial y_1},\quad
\frac{\partial u_2}{\partial y_1} =-\frac{\partial v_2}{\partial x_1},\quad
\frac{\partial u_2}{\partial x_2} = \frac{\partial v_2}{\partial y_2},\quad
\frac{\partial u_2}{\partial y_2} =-\frac{\partial v_2}{\partial x_2},
\end{align}
and we would have a system of differential equations in real quantities:
\begin{align}
\dot{x}_1=u_1(x_1,x_2,y_1,y_2),\quad&\dot{y}_1=v_1(x_1,x_2,y_1,y_2),\\
\dot{x}_2=u_2(x_1,x_2,y_1,y_2),\quad&\dot{y}_2=v_2(x_1,x_2,y_1,y_2).
\end{align}
If we were able to reduce the number of equations to two (in complex variables), it would be a small step forward.
Let us see the linear case first, to get some experience.
The system of real equations is:
\begin{align*}
\dot{x}_1&=ax_1+bx_2+cy_1+dy_2,\quad
&\dot{y}_1&=ex_1+fx_2+gy_1+hy_2,\\
\dot{x}_2&=Ax_1+Bx_2+Cy_1+Dy_2,\quad
&\dot{y}_2&=Ex_1+Fx_2+Gy_1+Hy_2,
\end{align*}
and the Cauchy--Riemann equations imply
\begin{equation*}
a=g, c=-e, b=h, d=-f,\quad
A=G, C=-E, B=H, D=-F.
\end{equation*}
The complex equations are:
\begin{equation*}
\dot{z}_1=(a-ci)z_1+(b-di)z_2,\quad
\dot{z}_2=(A-Ci)z_1+(B-Di)z_2.
\end{equation*}
Promising. But not from the point of reaction kinetics. Why?
To have a kinetic equation one should have
\begin{equation*}
c=d=e=f=C=D=E=F=0,    
\end{equation*}
thus the system falls apart:
\begin{align*}
\dot{x}_1&=ax_1+bx_2, &\dot{x}_2&=Ax_1+Bx_2,\\
\dot{y}_1&=gy_1+hy_2, &\dot{y}_2&=Gy_1+Hy_2,
\end{align*}

In the quadratic case, we start from
\begin{align}
\dot{x}_1
&=j_1x_1^2+j_2x_2^2+j_3y_1^2+j_4y_2^2\nonumber\\
&+j_5x_1x_2+j_6y_1y_2+j_7x_1y_1+j_8x_1y_2+j_9x_2y_1+j_{10}x_2y_2\nonumber\\ 
&+j_{11}x_1+j_{12}x_2+j_{13}y_1+j_{14}y_2+j_{15},\\
\dot{y}_1
&=k_1x_1^2+k_2x_2^2+k_3y_1^2+k_4y_2^2\nonumber\\
&+k_5x_1x_2+k_6y_1y_2+k_7x_1y_1+k_8x_1y_2+k_9x_2y_1+k_{10}x_2y_2\nonumber\\ 
&+k_{11}x_1+k_{12}x_2+k_{13}y_1+k_{14}y_2+k_{15},\\
\dot{x}_2
&=J_1x_1^2+J_2x_2^2+J_3y_1^2+J_4y_2^2\nonumber\\
&+J_5x_1x_2+J_6y_1y_2+J_7x_1y_1+J_8x_1y_2+J_9x_2y_1+J_{10}x_2y_2\nonumber \\ 
&+J_{11}x_1+J_{12}x_2+J_{13}y_1+J_{14}y_2+J_{15},\\
\dot{y}_2
&=K_1x_1^2+K_2x_2^2+K_3y_1^2+K_4y_2^2\nonumber\\
&+K_5x_1x_2+K_6y_1y_2+K_7x_1y_1+K_8x_1y_2+K_9x_2y_1+K_{10}x_2y_2\nonumber \\ 
&\quad+K_{11}x_1+K_{12}x_2+K_{13}y_1+K_{14}y_2+K_{15}.
\end{align}
and the Cauchy--Riemann equations imply
\begin{align*}
&k_1=-k_3=-\frac{j_7}{2},&  &k_2=-k_4=-\frac{j_{10}}{2}, &  &-k_5=k_6=j_8=j_9,\\
&k_7=2j_1=-2j_3,         &  &k_8=k_9=j_5=-j_6,           &  &k_{10}=2j_2=-2j_4,\\ &k_{11}=-j_{13},         &  &k_{12}=-j_{14}, &  &k_{13}=j_{11},\\
&k_{14}=j_{12},&&\\
&K_1=-K_3=-\frac{J_7}{2},&  &K_2=-K_4=-\frac{J_{10}}{2}, &  &K_5=-K_6=-J_8=-J_9,\\
&K_7=2J_1=-2J_3,         &  &K_8=K_9=J_5=-J_6,           &  &K_{10}=2J_2=-2J_4,\\
&K_{11}=-J_{13},         &  &K_{12}=-J_{14}, &  &K_{13}=J_{11},\\
&K_{14}=J_{12}.&&\\
\end{align*}
To use as few parameters as possible: 
altogether 2*11 (instead of 2*30), and these are \[j_1,j_2,j_5,j_7,j_8,j_{10},j_{11},j_{12},j_{13},j_{14},j_{15}\] and
\[J_1,J_2,J_5,J_7,J_8,J_{10},J_{11},J_{12},J_{13},J_{14},J_{15},\]
we shall use these relations.
\begin{align*}
&k_1=-\frac{j_7}{2},     &  &k_2=-\frac{j_{10}}{2},     &  &k_3=\frac{j_7}{2},\\
&k_4=\frac{j_{10}}{2},   &  &k_5=-j_8,                  &  &k_6=j_8,\\
&k_7=2j_1,               &  &k_8=j_5,                   &  &k_9=j_5,\\ 
&k_{10}=2j_2             &  &k_{11}=-j_{13},            &  &k_{12}=-j_{14},\\
&k_{13}=j_{11},          &  &k_{14}=j_{12},\\
&j_3=-j_1,               & &j_4=-j_2 &&j_6=-j_5&&j_9=j_8,\\
&K_1=-\frac{J_7}{2},     &  &K_2=-\frac{J_{10}}{2},     &  &K_3=\frac{J_7}{2},\\
&K_4=\frac{J_{10}}{2},   &  & K_5=-J_8,                 &  &K_6=J_8,\\
&K_7=2J_1,               &  &K_8=J_5                    &  &K_9=J_5, \\
&K_{10}=2J_2,            &  &K_{11}=-J_{13},            &  &K_{12}=-J_{14},\\
&K_{13}=J_{11},          &  &K_{14}=J_{12},\\
&J_3=-J_1,               & &J_4=-J_2 &&J_6=-J_5&&J_9=J_8.\\
\end{align*}

Thus, our system obeys the Cauchy--Riemann conditions are:
\begin{align}
\dot{x}_1
&=j_1x_1^2+j_2x_2^2-j_1y_1^2-j_2y_2^2\nonumber\\
&+j_5x_1x_2-j_5y_1y_2+j_7x_1y_1+j_8x_1y_2+j_8x_2y_1+j_{10}x_2y_2\nonumber\\ 
&+j_{11}x_1+j_{12}x_2+j_{13}y_1+j_{14}y_2+j_{15},\\
\dot{y}_1
&=-\frac{j_7}{2}x_1^2-\frac{j_{10}}{2}x_2^2+\frac{j_7}{2}y_1^2+\frac{j_{10}}{2}y_2^2\nonumber\\
&-j_8x_1x_2+j_8y_1y_2+2j_1x_1y_1+j_5x_1y_2+j_5x_2y_1+2j_2x_2y_2\nonumber\\ 
&-j_{13}x_1-j_{14}x_2+j_{11}y_1+j_{12}y_2+k_{15},\\
\dot{x}_2
&=J_1x_1^2+J_2x_2^2-J_1y_1^2-J_2y_2^2\nonumber\\
&+J_5x_1x_2-J_5y_1y_2+J_7x_1y_1+J_8x_1y_2+J_8x_2y_1+J_{10}x_2y_2\nonumber \\ 
&+J_{11}x_1+J_{12}x_2+J_{13}y_1+J_{14}y_2+J_{15},\\
\dot{y}_2
&=-\frac{J_7}{2}x_1^2-\frac{J_{10}}{2}x_2^2+\frac{J_7}{2}y_1^2+\frac{J_{10}}{2}y_2^2\nonumber\\
&-J_8x_1x_2+J_8y_1y_2+2J_1x_1y_1+J_5x_1y_2+J_5x_2y_1+2J_2x_2y_2\nonumber \\ 
&-J_{13}x_1-J_{14}x_2+J_{11}y_1+J_{12}y_2+K_{15}.
\end{align}
Therefore we have simpler complex equations as follows:
\begin{align}
    \dot{z}_1 &= (j_1 - i\frac{j_7}{2})z_1^2 + (j_2 - i\frac{j_{10}}{2})z_2^2 + (j_5 - ij_8)z_1z_2 + (j_{11} - ij_{13})z_1 \nonumber \\
    &\quad + (j_{12} - ij_{14})z_2 + (j_{15} + ik_{15}), \\
    \dot{z}_2 &= (J_1 - i\frac{J_7}{2})z_1^2 + (J_2 - i\frac{J_{10}}{2})z_2^2 + (J_5 - iJ_8)z_1z_2 + (J_{11} - iJ_{13})z_1 \nonumber \\
    &\quad + (J_{12} - iJ_{14})z_2 + (J_{15} + iK_{15}).
\end{align}
Assuming also that our system is kinetic, we have
\begin{align}
j_1,j_7\le0; j_2=j_5=j_8=j_{10}=j_{13}=j_{14}=0;j_{12},j_{15},k_{15}\ge0\\    
J_2,J_{10}\le0; J_1=J_5=J_7=J_8=J_{13}=J_{14}=0;J_{11},J_{15},K_{15}\ge0   
\end{align}
\begin{align}
\dot{x}_1
&=j_1x_1^2-j_1y_1^2+j_7x_1y_1+j_{11}x_1+j_{12}x_2+j_{15},\\
\dot{y}_1
&=-\frac{j_7}{2}x_1^2+\frac{j_7}{2}y_1^2
+2j_1x_1y_1+j_{11}y_1+j_{12}y_2+k_{15},\\
\dot{x}_2
&=J_2x_2^2-J_2y_2^2+J_{10}x_2y_2+J_{11}x_1+J_{12}x_2+J_{15},\\
\dot{y}_2
&=-\frac{J_{10}}{2}x_2^2+\frac{J_{10}}{2}y_2^2+2J_2x_2y_2+J_{11}y_1+J_{12}y_2+K_{15}.
\end{align}
The complex equations will also be simpler:
\begin{align}
    \dot{z}_1 &= (j_1 - i\frac{j_7}{2})z_1^2+j_{11}z_1+j_{12}z_2 + (j_{15} + ik_{15}), \\
    \dot{z}_2 &= (J_1 - i\frac{J_7}{2})z_2^2 +J_{11} z_1+J_{12}z_2 + (J_{15} + iK_{15}).
\end{align}
Here, we have reduced the four--variable real system into a two--variable complex system. In general, given a \(2n\)--variable real system, where \(n\) is a positive integer, we can reduce it to a \(n\)-variable complex system using the technique above.

If all the right-hand sides of the reduced system are homogeneous polynomials of the same degree then a further (final) step of reduction is also possible. \label{page:homog}
We show how to start the calculations with a simple example.
Suppose that we have arrived at two equations with homogeneous second-degree polynomials:
\begin{equation}
    \dot{z}_1 = a z_1^2 +  b z_1z_2 + c z_2^2 \quad
    \dot{z}_2 = A z_1^2 +  B z_1z_2 + C z_2^2. \label{eq:homsecond}
\end{equation}
Assuming that there is a function \(\Zeta\) such that \(z_2=\Zeta\circ z_1,\)
we have \(\dot{z_2}=\Zeta'\circ z_1 \dot{z_1},\) and
\begin{equation}
\Zeta'\circ z_1
=\frac{\dot{z_2}}{\dot{z_1}}
=\frac{A +  B \frac{\Zeta\circ z_1}{z_1} + C (\frac{\Zeta\circ z_1}{z_1})^2}{a +  b\frac{\Zeta\circ z_1}{z_1} + c (\frac{\Zeta\circ z_1}{z_1})^2}
\end{equation}
Let \(U(\zeta):=\frac{\Zeta(\zeta)}{\zeta},\) then \(\Zeta(\zeta)=\zeta U(\zeta),\) therefore
\({\Zeta'}(\zeta)=\zeta {U'}(\zeta)+U(\zeta),\) thus
\begin{equation}
\zeta {U'}(\zeta)+U(\zeta)
=\frac{A +  B U(\zeta) + C (U(\zeta))^2}{a +  bU(\zeta) + c (U(\zeta))^2},
\end{equation}
or
\begin{equation}
\zeta {U'}(\zeta)
=\frac{A +  B U(\zeta) + C (U(\zeta))^2}{a +  bU(\zeta) + c (U(\zeta))^2}-U(\zeta),
\end{equation}
and
\begin{equation}
{U'}(\zeta )\frac{a +  bU(\zeta) + c (U(\zeta))^2}{A+(B-a)U(\zeta)+(C-b)U(\zeta)^2-cU(\zeta)^3}=\frac{1}{\zeta}
\end{equation}
which is a separable differential equation for the function \(U.\) 
Solving via the Cardano formula one gets a relationship  \(\Phi(z_1(t),z_2(t))=0,\)
(which is a first integral of the equation \eqref{eq:homsecond})
that does not seem to be easily rearranged for the real and imaginary parts of the two functions \(z_1\) and \(z_2.\)
%%%%%%%%%%%%%%%%%%%%%%%%%%%%%%%%%%%%%%%%%%%
\section{Discussion and Outlook}
%%%%%%%%%%%%%%%%%%%%%%%%%%%%%%%%%%%%%%%%%%%%%%%
%%%%%%%%%%%%%%%%%%%%%%%%%%%%%%%%%%%%%%%%%%%%%%%
\subsection{Further directions}
%%%%%%%%%%%%%%%%%%%%%%%%%%%%%%%%%%%%%%%%%%%%%%%
%%%%%%%%%%%%%%%%%%%%%%%%%%%%%%%%%%%%%%%%%%%%%%%
\subsubsection{Partial differential equations}
%%%%%%%%%%%%%%%%%%%%%%%%%%%%%%%%%%%%%%%%%%%%%%%
Consider the system of reaction-diffusion equations
\begin{align}
\frac{\partial x(t,\rb)}{\partial t}=u(x(t,\rb),y(t,\rb))+D_x\Delta x(t,\rb)\\    
\frac{\partial y(t,\rb)}{\partial t}=v(x(t,\rb),y(t,\rb))+D_y\Delta y(t,\rb), 
\end{align}
Assuming \eqref{eq:cauchy}, and \(D_x=D_y,\) there exists \(f:=u+iv,\) furthermore one one gets a single, complex reaction-diffusion equation for the function \(z:=x+i y.\)
\begin{equation}
\frac{\partial z(t,\rb)}{\partial t}=f(z(t,\rb))+D\Delta z(t,\rb). 
\end{equation}
Note that the restriction that the two diffusion constants have the same value \(D\) is not too stringent: in aqueous solutions, this is approximately true for many materials.
%%%%%%%%%%%%%%%%%%%%%%%%%%%%%%%%%%%%%%%%%%%%%%%
\subsubsection{Approximations}
%%%%%%%%%%%%%%%%%%%%%%%%%%%%%%%%%%%%%%%%%%%%%%%
This example will also serve as an example of the second-degree case. Here we carry out all the calculations for a second-degree case with specified (numerical)
coefficients. One could substitute the concrete values of the parameters into the general formula, however---as in general when solving differential equations---it may be better to reproduce the derivation of the results in the concrete case.

\begin{figure}[ht]
  \centering
  \includegraphics[width=0.8\linewidth]{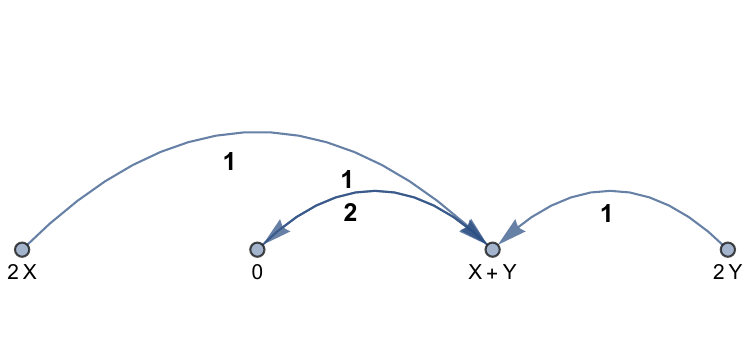}
  \caption{FHJ graph of the reaction \eqref{eq:secondex}}
  \label{fig:secondex}
\end{figure}

Fig. \ref{fig:secondex} shows the FHJ graph of the reaction \eqref{eq:secondex}.
\begin{equation}
\ce{2 X ->[1] X + Y <->[2][1] 0},\quad\ce{2 Y ->[1] X + Y} 
\label{eq:secondex}
\end{equation}

The induced kinetic differential equation of the reaction \eqref{eq:secondex}
with the initial concentrations \(x(0)=2\) and \(y(0)=1\) is
\begin{align}
&\dot{x}=1-x^2-2xy+y^2,\quad
\dot{y}=1+x^2-2xy-y^2, \label{eq:second01}   \\
&x(0)=2,\quad y(0)=1\nonumber
\end{align}
leading to the complex differential equation
\begin{equation}
\dot{z}=1+i+(-1+i)z^2,\quad z(0)=2+i.\label{eq:seconcomplex01}
\end{equation}

The solution of Eq. \eqref{eq:seconcomplex01} is
\begin{align*}
z(t)=\frac{(1+i) (-2 i e^{2 \sqrt{2} t}+3 (1+\sqrt{2}) e^{4 \sqrt{2} t}+3-3
   \sqrt{2})}{-8 e^{2 \sqrt{2} t}+3 (2+\sqrt{2}) e^{4 \sqrt{2} t}+6-3
   \sqrt{2}}\quad (t\in\R). 
\end{align*}
Therefore 
\begin{align}
x(t)&=\frac{2 e^{2 \sqrt{2} t}+3 (1+\sqrt{2}) e^{4 \sqrt{2} t}+3-3 \sqrt{2}}{-8 e^{2
   \sqrt{2} t}+3 (2+\sqrt{2}) e^{4 \sqrt{2} t}+6-3 \sqrt{2}}
 \quad (t\in\R), \label{eq:nicesol1}   \\ 
y(t)&=\frac{-2 e^{2 \sqrt{2} t}+3 (1+\sqrt{2}) e^{4 \sqrt{2} t}+3-3 \sqrt{2}}{-8 e^{2
   \sqrt{2} t}+3 (2+\sqrt{2}) e^{4 \sqrt{2} t}+6-3 \sqrt{2}}\quad (t\in\R). \label{eq:nicesol2}   
\end{align}
Note that the denominator in the above equations is never zero \ref{prop:nonzero}.

Now consider the equation 
\begin{align}
&\dot{x}=1-(1+\varepsilon)x^2-2xy+y^2,\quad
\dot{y}=1+x^2-2xy-y^2,  \label{eq:epsilon} \\
&x(0)=2,\quad y(0)=1\nonumber
\end{align}
While Eq. \eqref{eq:epsilon} is an equation fulfilling the Cauchy--Riemann condition for \(\varepsilon=0,\)
in general, it is not.
Note that it is also a kinetic differential equation for any real \(\varepsilon\).
As one has the explicit solution of the unperturbed equation one might hope that it is a good approximation of the solution to the perturbed equation.
\begin{figure}[ht]
  \begin{minipage}{0.5\textwidth}
    \centering
    \includegraphics[width=\linewidth]{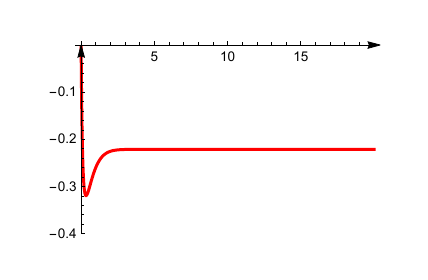}
    \caption{Differences between the first coordinates of the solution of \eqref{eq:epsilon} with \(\varepsilon=1\) and \(\varepsilon=0\).}
    \label{fig:xdifferences}
  \end{minipage}\quad
  \begin{minipage}{0.5\textwidth}
    \centering
    \includegraphics[width=\linewidth]{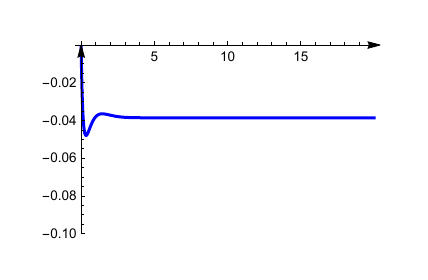}
    \caption{Differences between the second coordinates of the solution of \eqref{eq:epsilon} with \(\varepsilon=1\) and \(\varepsilon=0\).}
    \label{fig:ydifferences}
  \end{minipage}
\end{figure}

Figures \ref{fig:xdifferences} and \ref{fig:ydifferences} show the differences between the first, respectively, second coordinates of the solutions of \eqref{eq:epsilon} with \(\varepsilon=1\) and \(\varepsilon=0.\) Note that for  \(\varepsilon=0\) the solutions can be determined symbolically, as it is a Cauchy--Riemann--Erugin equation.
%%%%%%%%%%%%%%%%%%%%%%%%%%%%%%%%%%%%%%%%%%%%%%%
\subsection{Final remarks}
%%%%%%%%%%%%%%%%%%%%%%%%%%%%%%%%%%%%%%%%%%%%%%%
The closest mathematical paper to ours has been written by Calogero and Payandeh \cite{calogeropayandehcoupled}. 
Now, we will systematically compare the two works. 
Their main result can be summarized in the following statement.
\begin{theorem}
Consider the equations Eq. (1b)--Eq. (1c) of \cite{calogeropayandehcoupled}
\begin{align}
\dot{x_1}&=c_{11}x_{1}^2 + c_{12}x_1x_2+c_{13}x_{2}^2+c_{14}x_1+c_{15}x_2+c_{16},\label{eq:CP1}\\ 
\dot{x_2}&= c_{21}x_{1}^2+c_{22}x_1x_2+c_{23}x_{2}^2+c_{24}x_1+c_{25}x_2+c_{26},\label{eq:CP2}
\end{align}
and assume that the equalities
\begin{align}
&4c_{13}c_{21}-c_{12}c_{22}=0                                       \label{eq:CPequ1}\\
&2(-c_{12}+2c_{23})c_{21} + (2c_{11}-c_{22})c_{22} = 0 \label{eq:CPequ2}\\
&c_{24}(2c_{11}-c_{22}) +2c_{21} (c_{25}-c_{14}) = 0     \label{eq:CPequ3}\\
&c_{12}c_{24} - 2c_{15}c_{21} = 0                                   \label{eq:CPequ4}
\end{align}
 hold. 
Furthermore, assume that none of the denominators in the steps
(39a)--(44f) in \cite{calogeropayandehcoupled}, describing the algorithm, are different from zero. 
Then, the explicit symbolic solution of \eqref{eq:CP1}--\eqref{eq:CP2} can be constructed by applying the steps of the algorithm  (39a)--(44f)   in \cite{calogeropayandehcoupled}.
\end{theorem}
If  \eqref{eq:CP1}--\eqref{eq:CP2} is a Cauchy--Riemann--Erugin system, then the coefficients fulfill \eqref{eq:CPequ1}--\eqref{eq:CPequ4}. 

To show that this method implies our method under some conditions, 
we try to obtain the above constraints by assuming that the right-hand side of 
\eqref{eq:CP1}--\eqref{eq:CP2} satisfy Cauchy--Riemann equations:
\begin{align*}
2c_{11} - c_{22} =0, \quad 
2c_{23}-c_{12} =0, \quad
c_{14} - c_{25}= 0, \quad
2c_{21} =-c_{12},  \quad
2c_{13}=-c_{22}, \quad
c_{15} = -c_{24}.
\end{align*}
Simple substitutions show the statement.

It is important to note that this does not imply that all the Cauchy--Riemann--Erugin systems can be solved by the method of 
Calogero and Payandeh; for example, the linear systems cannot, because they do not satisfy the conditions about the denominators.  

In the following cases, we demonstrate that the far-reaching method presented by Calogero and Payandeh may also lead to a dead end, awaiting further simplification methods.

In the example 
\begin{align}
\dot{x}_1&=1 + 2 x + 2 x^2 + y + 2 x y + 2 y^2, \label{eq:ex1}\\
\dot{y}&= 1 + x + x^2 + 2 y + 4 x y + y^2 \label{eq:ex2}
\end{align}
the Cauchy--Riemann conditions are not fulfilled, but the Calogero--Payandeh conditions are, including the fact that none of the corresponding denominators are different from zero,therefore this equation can be solved by their method, and cannot be solved by ours. What is more, there are no "hidden" \(i\) numbers under the square roots:
\begin{align}
x(t)&=\frac{57 (3 e^t-2)+\sqrt{15} (2 e^t-25) \sin (\sqrt{15} t)-3 (62 e^t-53) \cos
   (\sqrt{15} t)}{3 (e^t-2) (7 \sqrt{15} \sin (\sqrt{15} t)+71 \cos (\sqrt{15}
   t)-76)},\\
y(t)&=\frac{-57 (e^t+2)+\sqrt{15} (23 e^t-25) \sin (\sqrt{15} t)+3 (9 e^t+53)
   \cos (\sqrt{15} t)}{3 (e^t-2) (7 \sqrt{15} \sin (\sqrt{15} t)+71 \cos (\sqrt{15}
   t)-76)}.
\end{align}
The denominators in \cite{calogeropayandehcoupled} are usually more or less complicated expressions (polynomials) of the coefficients, except the denominator in Eq. (41) of \cite{calogeropayandehcoupled} that seems to be very hard to analyze.

In summary, there is a significant overlap in the scopes of the two methods concerning two-variable second-degree systems. 
However, we have also provided some ideas for cases outside this class of systems. 
If you truly need the symbolic solution of an equation of the form \eqref{eq:CP1}--\eqref{eq:CP2} you should try everything offered by the two papers.
Finally, one should not overlook the computational difficulties. 
For instance, in the case of \eqref{eq:second01}, we were able to arrive at an explicit solution \eqref{eq:nicesol1}--\eqref{eq:nicesol2}, but we struggled in vain to demonstrate its identity with the result obtained by the method of \cite{calogeropayandehcoupled}, despite attempting all the simplifying functions available in the Wolfram language.

We have seen that in the linear case, our method requires lengthier calculations than the traditional approach. However, in the second-degree case, our method is effective as opposed to the traditional one.

In \cite{calogeropayandehperiodic} the authors
can solve equations of specific form under restrictions on the parameters.
In one of their example, they find (explicitly) a periodic solution, 
and to arrive at this they use an equation for a complex-valued function 
and expand it into real and imaginary parts.
\cite{zhangzhang} go in the opposite direction: they use a five-variable real equation to prove the boundedness of the complex Lorenz system.

Our method can also be used to \emp{approximate} solutions of (kinetic or other) differential equations. If our polynomial equation is close to a Cauchy--Riemann--Erugin system, then one can solve that equation,
and use the Peano inequality to estimate the difference between the obtained solution and the solution we are interested in. If we are in the position to make the difference between the right-hand sides then we can make the difference of the solutions arbitrarily small \emp{on any finite interval}.

There is some hope that some special four-variable systems can be solved similarly by applying quaternionic calculus, e.g. 
\cite[Theorem 3.1]{deavours} might be useful for this purpose. See also \cite{gentilistoppatostruppa,sudbery}.
%%%%%%%%%%%%%%%%%%%%%%%%%%%%%%%%%%
\section{Appendix}
%%%%%%%%%%%%%%%%%%%%%%%%%%%%%%%%%%%
%%%%%%%%%%%%%%%%%%%%%%%%%%%%%%%%%%%%%%%%%%%%%%%%%%%%%%%%%%%%%%%%%%%%%%
\subsection{Useful formulas}
%%%%%%%%%%%%%%%%%%%%%%%%%%%%%%%%%%%%%%%%%%%%%%%%%%%%%%%%%%%%%%%%%%%%%%
Before anything else, we present two formulas we are recurrently using.
\begin{lemma}\label{lem:formulas}
\begin{enumerate}
\begin{comment}
Suppose \(\alpha,\beta,\gamma,\delta\) are real numbers so that \(\gamma^2+\delta^2\neq0.\) Then:
\begin{equation}
\frac{\alpha+\beta i}{\gamma+\delta i}=\ratio{\alpha}{\gamma}{\beta}{\delta}.\label{eq:ratio}
\end{equation}
\end{comment}
\item[\nonumber]
Suppose \(\xi\) and \(\eta\neq0\) are real numbers. 
Then the square roots of \(\xi+i\eta\) are as follows: 
\begin{align}
&\res{\xi}{\eta}\label{eq:squareroot1}\\
=&\result{\xi}{\eta}\label{eq:squareroot2}
\end{align}
\end{enumerate}
\end{lemma}
%%%%%%%%%%%%%%%%%%%%%%%%%%%%%%%%%%%%%%%%%%%%%%%%%%%%%%%%%%%%%%%%%%%%%%
\subsection{Addendum to the second-degree case}\label{subsec:addsecond}
%%%%%%%%%%%%%%%%%%%%%%%%%%%%%%%%%%%%%%%%%%%%%%%%%%%%%%%%%%%%%%%%%%%%%%
Let us start with the solution of \eqref{eq:secondcomplex}
\begin{equation}\label{eq:secondcomplexsol}
z(t)
=z_1+\frac{z_2-z_1}{1-\frac{z(0)-z_2}{z(0)-z_1}e^{(d-i\frac{e}{2})(z_2-z_1) t}}
=z_2+\frac{z_1-z_2}{1-\frac{z(0)-z_1}{z(0)-z_2}e^{(d-i\frac{e}{2})(z_1-z_2) t}}
\end{equation}
where \(z_1\) and \(z_2\) are the roots of the polynomial \(z\mapsto(a+iA)+(b-ic)z+(d-i\frac{e}{2})z^2,\)  and \(z(0):=x_0+iy_0.\)
To get an explicit solution, further---although elementary---lengthy calculations are needed.
Let us apply Euler's formula.
\begin{align}\label{eq:secondderivation}
&z(t)=\\
&z_1+\frac{z_2-z_1}{1-\frac{z(0)-z_2}{z(0)-z_1}e^{(d-i\frac{e}{2})(z_2-z_1) t}}=\\
&x_1+iy_1+\frac
{x_2-x_1+i(y_2-y_1)}
{1-\frac{x_0-x_2+i(y_0-y_2)}{x_0-x_1+i(y_0-y_1)}e^{(d(x_2-x_1)+\frac{e}{2}(y_2-y_1))t}\left(\cos(\Omega t)+i\sin(\Omega t)\right)}
\end{align}
with \(\Omega:=-\frac{e}{2}(x_2-x_1)+d(y_2-y_1),\) 
and with \(x_1+iy_1:=z_1, x_2+iy_2:=z_2.\)

\begin{proposition}
Assume that the real numbers \(a,b,c,d,e,A\) are such that \((d-i\frac{e}{2})\neq0.\) Then, the roots of the polynomial
\begin{equation}\label{eq:polysecond}
z\mapsto a+iA+(b-ic)z+(d-i\frac{e}{2})z^2
\end{equation}
are as follows
\begin{align}
z_{1,2}&=\frac{-b+ic\pm (q+ir)}{2d-ie}=
\frac{-b\pm q+i(c\pm r)}{2d-ie}\label{eq:root01}\\
&=\left(\frac{\pm 2dq-2bd-ec\mp er}{4d^2+e^2}+i\frac{2cd\pm 2dr\pm eq-be}
{4d^2+e^2}\right)\label{eq:root02}
\end{align}
with
\begin{align}
q&:=\sqrt{\frac{1}{2}(\dsqr+\sqrt{(\dsqr)^2+4(\deaF)^2})},\\
r&:=\frac{-\sqrt{2}(\deaF)}{\sqrt{\dsqr+\sqrt{(\dsqr)^2+4(\deaF)^2})}}.     
\end{align}
\end{proposition}
\Pf{
\begin{enumerate}
\item
To go from \eqref{eq:root01} to \eqref{eq:root02} apply the rule for ratios.
\item
Next, let us apply \eqref{eq:squareroot1} with \(\xi:={\dsqr},\quad \eta:={-2(\deaF)}\) to get
\eqref{eq:root02}.
\end{enumerate}
}
Using the explicit formula for the roots one can further expand the solution to \eqref{eq:secondcomplex}, but if one makes all the necessary substitutions with \(q\) and \(r\) as defined above, the results will be useless. However, we shall show examples that the procedure can be finished in all special cases.
%%%%%%%%%%%%%%%%%%%%%%%%%%%%%%%%%%%%%%%%%%%%%%%%%%%%%%%%%%%%%%%%%%%%%%
\subsection{Addendum to an example of the second-degree case}\label{subsec:seconddegree}
%%%%%%%%%%%%%%%%%%%%%%%%%%%%%%%%%%%%%%%%%%%%%%%%%%%%%%%%%%%%%%%%%%%%%%
\begin{proposition}\label{prop:nonzero}
    The denominator in the equations \eqref{eq:seconcomplex01} is never zero.
\end{proposition}
\Pf{
Let \(y:=e^{2\sqrt{2}t}\), then we have the following expression in the denominator.
\begin{equation}\label{eq:denom}
    3 (2+\sqrt{2})y^2-8y+(6-3 \sqrt{2})
\end{equation}
The zeros of \eqref{eq:denom} above are:
\begin{equation*}
    y_1 = \frac{4-i\sqrt{2}}{6+3\sqrt{2}}, \quad y_2=\frac{4+i\sqrt{2}}{6+3\sqrt{2}}
\end{equation*}
But \(y:=e^{2\sqrt{2}t}\) is always a real number for all \(t \in \Rb\) and cannot be equal to the
above complex roots.
Therefore, \(x(t)\) and \(y(t)\) are always defined for all real numbers \(t \in \Rb\).
}
%%%%%%%%%%%%%%%%%%%%%%%%%%%%%%
%%%%%%%%%%%%%%%%%%%%%%%%%%%%%%%%
%%%%%%%%%%%%%%%%%%%%%%%%%%%%%%%%%%%%%%%%%%%%%%%%%%%%%%%%%%%%%%%%%%%%%%
\subsection{Addendum to the third-degree case}\label{subsec:thirddegreecase}
%%%%%%%%%%%%%%%%%%%%%%%%%%%%%%%%%%%%%%%%%%%%%%%%%%%%%%%%%%%%%%%%%%%%%%

\janos{FHJ graphs}

Let us apply the proposed procedure in some typical cases.
\begin{enumerate}
\item 
The real initial value problem
\begin{align}
&\dot{x}=-8+12x-6x^2+6y^2+x^3-3xy^2,\quad
\dot{y}=12y-12xy+3x^2y-y^3, \label{eq:third01}   \\
&x(0)=2,\quad y(0)=1\nonumber
\end{align}
leads to the complex differential equation
\begin{equation*}
\dot{z}= (z-2)^3,\quad z(0)=2+i
\end{equation*}
having the solution 
\begin{equation}
z(t)=2+\frac{i}{\sqrt{1+2t}}, \quad (t\in]-\frac{1}{2},+\infty[),    
\end{equation}
or
\begin{equation}
x(t)=2,\quad y(t)=\frac{1}{\sqrt{1+2t}}, \quad (t\in]-\frac{1}{2},+\infty[).    
\end{equation}
Substituting this result into \eqref{eq:third01} verifies the calculations.
\item 
One can proceed similarly if the starting initial value problem is
\begin{align}
&\dot{x}=-3x+6xy+x^3-3xy^2,\quad
\dot{y}=1-3y-3x^2+3y^2+3x^2y-y^3,\label{eq:third02}\\
&x(0)=1,\quad y(0)=2\nonumber.
\end{align}
The complex differential equation is now:
\begin{equation*}
\dot{z}= (z-i)^3,\quad z(0)=1+2i
\end{equation*}
having the solution
\begin{align*}
z(t)&=i+\frac{2}{\sqrt{1-8t}}, \quad (t\in]-\infty,\frac{1}{8}[),  
\end{align*}
with
\begin{equation}
x(t)=\frac{2}{\sqrt{1-8t}},\quad (t\in]-\infty,\frac{1}{8}[),  
y(t)=1, \quad (t\in\R)    
\end{equation}
\end{enumerate}
As can be seen from the above examples, if the right-hand side of the complex equation has a triple root, the solutions of the real system can be expressed symbolically.
One can never have this nice form with kinetic differential equations.
\begin{proposition}
The right-hand side $(z-a-ib)^3; \quad a,b\in\R; a^2+b^2\neq0$ always comes from a real system containing negative cross-effect(s). 
\end{proposition}
\Pf{Since 
\begin{align*}
(z-a-ib)^3
=&-a^3+3 a b^2+3(a^2-b^2) x-6 a b y-3 a x^2+6 b x y+3 a y^2+x^3-3 x y^2\\
+&i(-3 a^2 b+b^3+6 a b x+3 (a^2 - b^2) y-3 b x^2-6 a x y+3 b y^2+3 x^2 y-y^3)
\end{align*}
\(6ab\le0\) and \(6ab\ge0\) implies that \(6ab=0.\)
If \(a=0,\) then \(b^3\ge0\) and \(-3b\ge0\) should hold, thus \(b=0\) would be implied.
If \(b=0,\) then \(-a^3\ge0\) and \(3a\ge0\) should hold, thus \(a=0\) would be implied.}
\begin{remark}
In other cases,  one may either have terrible formulas based on the Cardano formula,
or one may be content with an implicit solution for the real components.
E.g. the system
\begin{align}
\dot{x}&=1-3y-2x^2+4xy+2y^2+x^3-3xy^2,\\
\dot{y}&=-1+3x-2x^2-4xy+2y^2+3x^2y-y^3
\end{align}
leads to the complex differential equation:
\begin{equation}
\dot{z}= (z-1)(z-i)(z-i-1),\quad z(0)=x_0+iy_0,
\end{equation}
the solution of which can either be obtained using the Cardano formula,
but the result is too complicated to be useful,
or it can be left in the form to produce an implicit solution, first for \(z(t),\)
then for \(x(t), y(t).\)
\end{remark}

%%%%%%%%%%%%%%%%%%%%%%%%%%%%%%%%%%%%%%%%%%%%%%%%%%%%%%%%%%%%%%%%%%%%%%
\subsection{An example of the four-variable case}
%%%%%%%%%%%%%%%%%%%%%%%%%%%%%%%%%%%%%%%%%%%%%%%%%%%%%%%%%%%%%%%%%%%%%%
Let us start with a system of equations with a homogeneous right-hand side.
\begin{align}
\dot{x}_1&=ax_1^2+bx_2^2+cy_1^2+dy_2^2+ex_1x_2+fy_1y_2+gx_1y_1+hx_1y_2+ix_2y_1+jx_2y_2,\\
\dot{y}_1&=kx_1^2+lx_2^2+my_1^2+ny_2^2+px_1x_2+qy_1y_2+rx_1y_1+sx_1y_2+tx_2y_1+ux_2y_2,\\
\dot{x}_2&=Ax_1^2+Bx_2^2+Cy_1^2+Dy_2^2+Ex_1x_2+Fy_1y_2+Gx_1y_1+Hx_1y_2+Ix_2y_1+Jx_2y_2,\\
\dot{y}_2&=Kx_1^2+Lx_2^2+My_1^2+Ny_2^2+Px_1x_2+Qy_1y_2+Rx_1y_1+Sx_1y_2+Tx_2y_1+Ux_2y_2.
\end{align}
and the Cauchy--Riemann equations imply
\begin{align}
&a=-c=\frac{r}{2},e=t=s=-f,h=q=i=-p,\frac{g}{2}=m=-k,b=-d=\frac{u}{2},n=-l=\frac{j}{2},\\
&A=-C=\frac{R}{2},E=T=S=-F,H=Q=I=-P,\frac{G}{2}=M=-K,B=-D=\frac{U}{2},N=-L=\frac{J}{2}.
\end{align}
therefore we have simpler complex equations as follows:
\begin{align}
    \dot{z}_1 &= (a-i\frac{g}{2})z_1^2 + (b-i\frac{j}{2})z_2^2+(e-ih)z_1z_2\\
    \dot{z}_2 &= (A-i\frac{G}{2})z_1^2 + (B-i\frac{J}{2})z_2^2+(E-iH)z_1z_2
\end{align}
Now we can again use the method described in \pageref{page:homog}.
\bibliographystyle{plain}
\bibliography{_KipronoProblems} 
%\nocite{*}
\end{document}